**Phase transitions of eutectic high entropy alloy $AlCoCrFeNi_{2.1}$ under shock compression**

Sophie Emilene Parsons[1,2,3], Kento Katagiri[1,2,3*], Hangman Chen[4], Anirudh Hari[1,2,3], Sara Jessica Irvine[2,3,5], Dorian Luccioni[1,2,3], Rayen Lin[1,3,6], Laura Elizabeth Madril[1,2,3], Tharun Reddy[1,2,3], William J. McKinney[4], Jie Ren[7], Wuxian Yang[8], Norimasa Ozaki[9,10], Alexis Amouretti[9], Ryosuke Kodama[9,10], Hirotaka Nakamura[9], Yusuke Nakanishi[9], Masato Ota[11], Yusuke Seto[12], Sota Takagi[13], Takuo Okuchi[14], Yuhei Umeda[14], Yuichi Inubushi[15,16], Kohei Miyanishi[16], Keiichi Sueda[16], Tadashi Togashi[15,16], Makina Yabashi[15,16], Toshinori Yabuuchi[15,16], Wanghui Li[17], Paul Elliot Specht[18], Penghui Cao[4], Wen Chen[8], Yogesh K. Vohra[19], Leora Dresselhaus-Marais[1,2,3,*]

[1]Department of Materials Science and Engineering, Stanford University, California, USA.

[2]SLAC National Accelerator Laboratory, California, USA.

[3]PULSE Institute, Stanford University, California, USA.

[4]Department of Mechanical and Aerospace Engineering, University of California, Irvine, California, USA.

[5]Department of Applied Physics, Stanford University, California, USA.

[6]Department of Mechanical Engineering, University of California, Los Angeles, California, USA.

[7]Department of Mechanical and Industrial Engineering, University of Massachusetts, Massachusetts, USA.

[8]Department of Aerospace and Mechanical Engineering, University of Southern California, California, USA.

[9]Graduate School of Engineering, Osaka University, Osaka, Japan.

[10]Institute of Laser Engineering, Osaka University, Osaka, Japan.

[11]National Institute for Fusion Science, Gifu, Japan.

[12]Graduate School of Science, Osaka Metropolitan University, Osaka, Japan.

[13]Earth and Planets Laboratory, Carnegie Institution for Science, Washington DC, USA.

[14]Institute for Integrated Radiation and Nuclear Science, Kyoto University, Osaka, Japan.

[15]Japan Synchrotron Radiation Research Institute, Hyogo, Japan.

[16]RIKEN SPring-8 Center, Hyogo, Japan.

[17]Institute of High Performance Computing (IHPC), Agency for Science, Technology and Research (A*STAR), Singapore 138632, Republic of Singapore

[18]Sandia National Laboratories, Albuquerque, New Mexico, USA.

[19]Department of Physics, University of Alabama at Birmingham, Alabama, USA.

*Corresponding authors. Email: kentok@stanford.edu & leoradm@stanford.edu

**High entropy alloys (HEAs) are a new class of metals that exhibit unique mechanical performance. Among HEAs, additively manufactured eutectic high entropy alloys (AM-EHEAs) have recently emerged as candidate materials for use in extreme conditions due to their simultaneous high strength and ductility. However, the deformation and structural evolution of AM-EHEAs under conditions of high pressure have not been well characterized, limiting their use in extreme applications. We present dynamic compression experiments and molecular dynamics simulations studying the structural evolution of AM-EHEA $AlCoCrFeNi_{2.1}$ when compressed to pressures up to 400 GPa. Our *in-situ* X-ray diffraction measurements capture the appearance of fcc and bcc phases at different pressure conditions, with pure- and mixed-phase regions. Understanding the phase stability and structural evolution of the AM EHEA offers new insights to guide the development of high-performance complex materials for extreme conditions.**

## I. Introduction

The design of alloys with high strength and favorable mechanical properties has long been critical to structural materials for construction, automotive, and aerospace industries. For instance, in the design of building materials resilient to natural disasters, materials that retain strength to high strain rates and exhibit ductility are necessary for structure survival[1]. Traditionally, alloys have consisted of one major element with other elements used in minor amounts to tune properties. One of the biggest challenges of alloy design for extreme applications is that the strongest materials (e.g. diamond) are typically brittle and fracture instead of deforming plastically upon loading. High entropy alloys (HEAs), first demonstrated in 2004, include multiple elements as principal components in nearly equimolar ratios[2]. The entropic stabilization proposed to be the rationale for their inherent stability disrupts classical trade-offs in alloy design like the counter-acting trends between strength vs ductility. Today HEAs are transforming alloy design, as some HEAs exhibit remarkable hardness, high-temperature strength, fatigue resistance, and corrosion resistance[3]. For conditions that typically struggle to find appropriate alloys stable at their thermomechanical conditions, the high-pressure stability proposed for HEAs are ideal for those applications[4–6].

In steel, phase stability has long been linked to the strength and failure behavior of alloys under static and dynamic compression[7]. Under quasi-static loading, steels like austenitic stainless steel can transform to martensite, increasing work hardening and strength. However, higher strain rates reduce the amount of martensite formed, lessening this strengthening effect[8]. Some steel alloys further show a strain rate dependent response. At low rates relevant to quasi-static experiments these alloys demonstrate little variation in material response but show a pronounced increase in strength at higher rates, often linked to phase transformation and microstructural changes[9]. A transition from slip to twinning has been shown to be activated due to phase changes and strain rate increase, demonstrating the importance of understanding high strain rate behavior of these alloys[10].

The same is also true for HEAs. Traditional HEAs have shown that the deformation behavior is influenced strongly by the strain rate of the loading conditions. To date, there are no dynamic studies of dual phased HEAs far from equilibrium up to the melt. Understanding the complex phase diagram of HEAs is essential to understand phase boundary strengthening in this unusual class of materials. Quantifying the phase diagram is essential to define the interplay between phase transitions and plasticity at extremes. Clarifying the deformation behavior of this class of material allows for the future fabrication of HEAs with favorable strength qualities to higher pressures relevant to defense and aerospace applications. Mapping the high strain rate behavior of this HEAs also increases the predictive ability of simulations, providing much needed data to improve potentials to higher pressure states.

The state of the art in mechanical properties of HEAs is dual-phase HEAs due to their strong yet ductile behavior. Dual-phase HEAs have been developed to uniquely enhance strength and ductility commensurately by introducing phase boundaries between soft and hard phases. In particular, eutectic HEAs (EHEAs) have been used to engineer the grain structures using cooling rate without changing the composition of the eutectic colonies. EHEAs are composed of eutectic colonies with at least two solid phases that form simultaneously during solidification. For example, $AlCoCrFeNi_{2.1}$ composition EHEAs form microlamellar structures of alternating soft face-centered cubic (fcc) and hard body-centered cubic (bcc) phases when cast, leading to simultaneous high strength and ductility[11]. The microlamellar structures that typically form in $AlCoCrFeNi_{2.1}$ have strengthening effects[12] that are amplified when 3D printed due to formation of nanolamellar structures that offer additional strengthening due to the Hall-Petch effect[13]. The resulting nanolamellar structure demonstrated unprecedented strength through nanoscale stress transfer without compromising the high ductility – the first material with these commensurate properties.

A few high-pressure loading studies have begun to explore the phase diagram of AM EHEA $AlCoCrFeNi_{2.1}$ under specific conditions. Previous measurements on AM $AlCoCrFeNi_{2.1}$ studied the quasi-static pressure dependence of the metal at room temperature, and a melting point measurement. When statically compressed at room temperature, the initial fcc/bcc eutectic structure transformed to fully fcc at 21 ± 3 GPa[14] and remains stable to 302 GPa[15]. The latter study reported that the AM EHEA melts at 1648 ± 25 K at 5.6 GPa. Large areas of phase space of AM EHEA $AlCoCrFeNi_{2.1}$ at high-pressure remain unexplored, leaving large uncertainties in the phase diagram of AM EHEA $AlCoCrFeNi_{2.1}$.

Mapping the phase diagram of the strong and ductile AM $AlCoCrFeNi_{2.1}$ is crucial to establish the operating conditions for which this alloy could exhibit phase-boundary strengthening in industry. In this study, we conducted dynamic and static compression experiments to map the phase diagram of AM $AlCoCrFeNi_{2.1}$. Using *in-situ* XRD measurements, we observed multiple phase transitions of the EHEA at high pressures and temperatures. We used the laser-shock technique to probe the highest pressures and temperatures along the equation of states defining shock compression (the Hugoniot). We used molecular dynamics (MD) simulations to predict the shock temperature. Leveraging dynamic compression experiments and MD simulations comprehensively maps the phase diagram of AM $AlCoCrFeNi_{2.1}$ from ambient conditions to 400 GPa.

## II. Results and Discussion

## A. Laser shock experiments

The laser shock experiments were conducted at Experimental Hutch 5 (EH-5) at SPring-8 Angstrom Compact Free Electron Laser (SACLA). At EH-5, the X-ray beam generated by the X-ray Free Electron Laser (XFEL)[16,17] is synchronized to a 532-nm nanosecond optical laser with 15-J pulses that generate shock waves in a laser-pump X-ray probe geometry (Fig 1a)[18]. A variable attenuator is used to adjust the laser energy arriving the target between 3 and 15 J, allowing us to tune the shock pressure in the sample. The nanosecond optical laser initiates a strong shock wave in the ablator material that transmits into the sample and reaches a high pressure and temperature state along the Hugoniot equation of state (EoS). The femtosecond-duration XFEL pulses are used to obtain structural information about the shocked sample via XRD. We monitored the shock breakout timing using line Velocity Interferometer System for Any Reflector (line-VISAR) and adjusted the XFEL timing to measure the shock compression before onset of release[19,20]. These timings correspond to probing the compressed material at times just before the wave breaks out of the back surface of the sample.

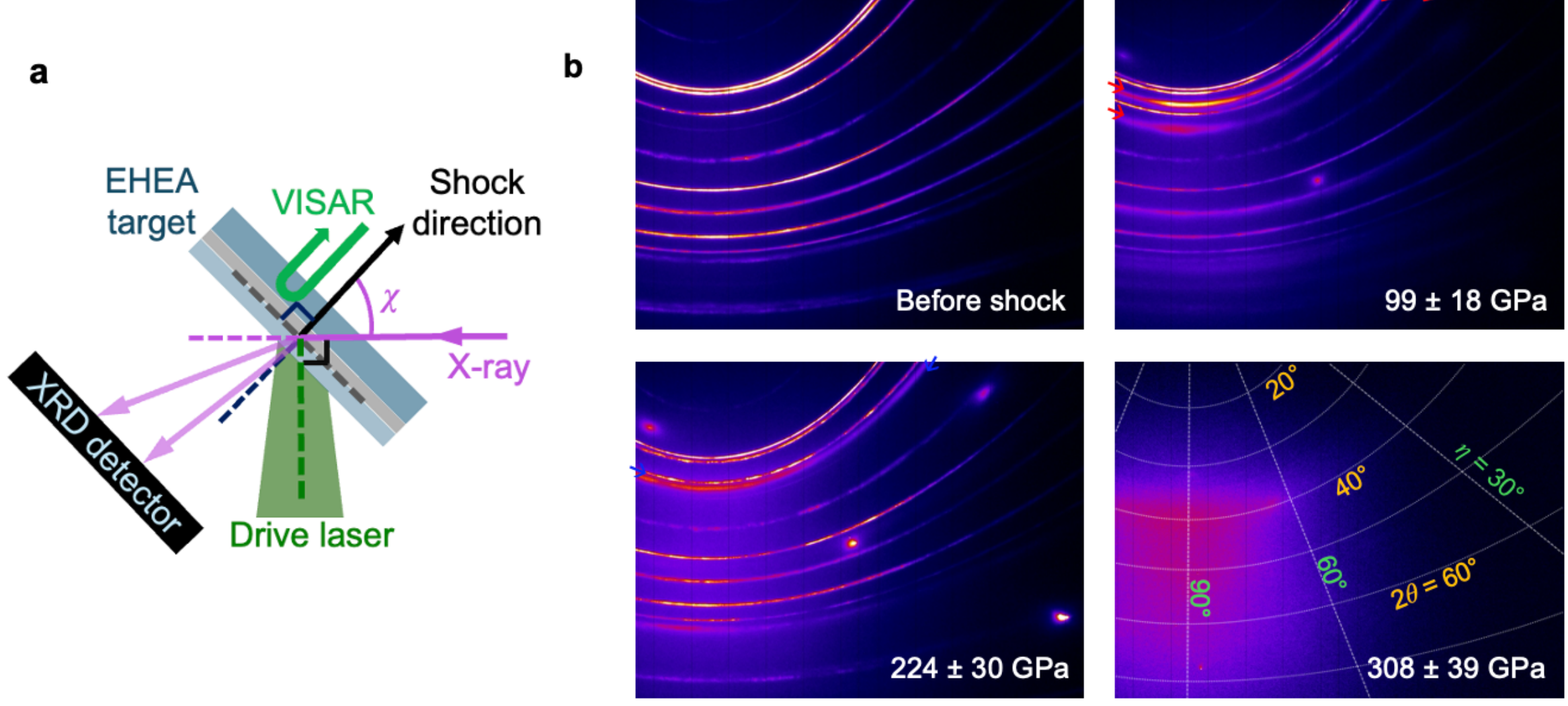

**Figure 1.** *In-situ* XRD measurements of AM $AlCoCrFeNi_{2.1}$ during shock compression, with blue indicators to denote bcc phase peaks and red indicators to denote fcc phase peaks. (**a**) Experimental diagram of the target, the drive laser, and the femtosecond XFEL pulse with simultaneous line-VISAR measurements. (**b**) Collected XRD of samples before shock compression, and during shocks to pressures of 99 ± 18, 224 ± 30, and 308 ± 39 GPa. Colored arrows indicate diffraction rings of interest; bright spots at 99 and 224 GPa are from uncompressed lithium fluoride windows, and weak spots at 308 GPa are from a quartz window. White lines and curvatures shown in the 308 ± 39 GPa image indicate constant azimuthal ($\eta$) and scattering ($2\theta$) angles, respectively.

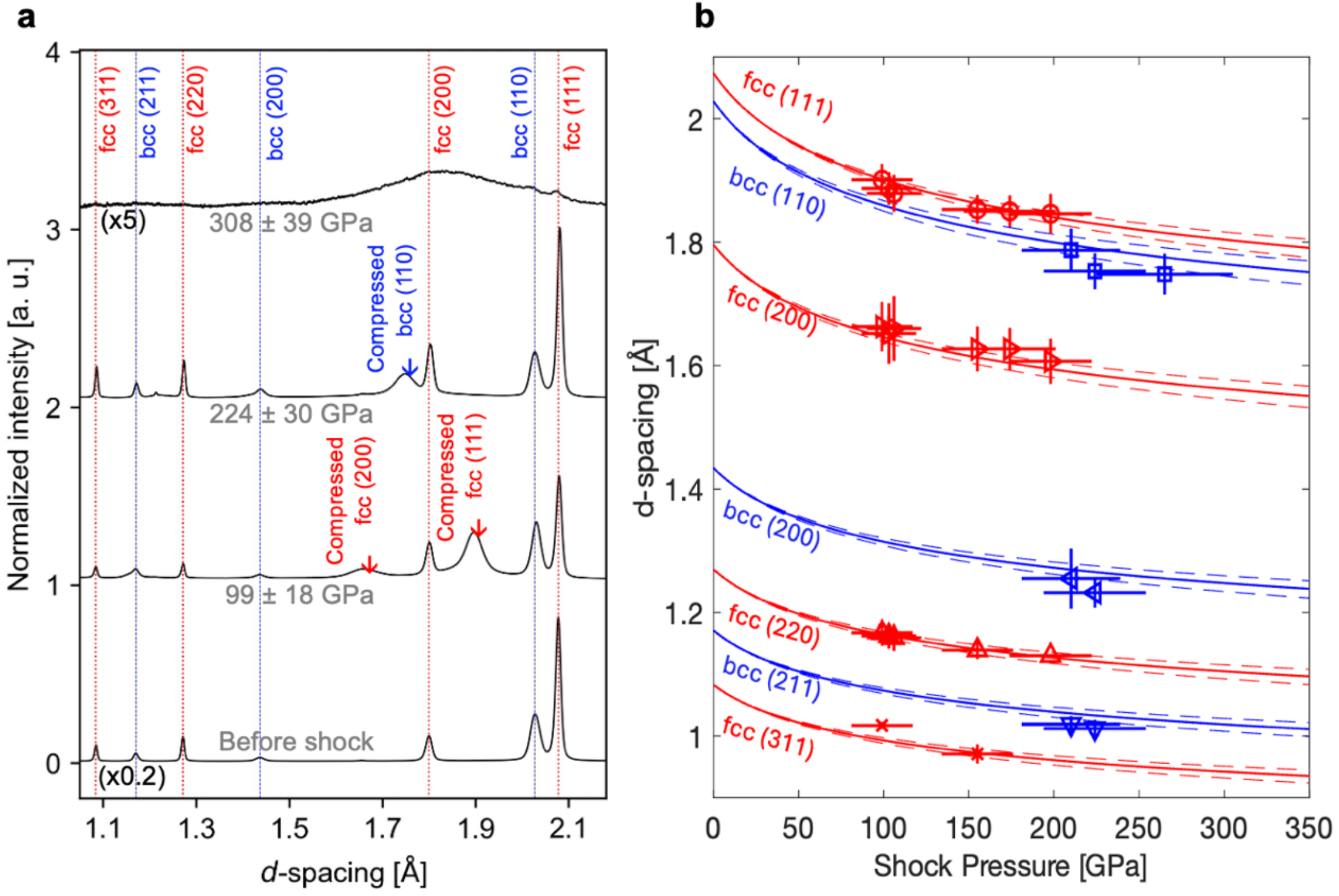

**Figure 2:** (**a**) Line profiles of the XRD images shown in Fig. 2b, unnormalized and offset for comparison. (**b**) Plot of the fitted *d*-spacings of the fcc and bcc phases as a function of shock pressure (dashed curves) with our data overlayed. Uncertainty in Hugoniot is represented by dashed lines.

By using XRD, we can experimentally observe the phase transformations undergone by this material when compressed to different pressures along its Hugoniot EoS. Fig. 1b shows example XRD patterns for different peak pressures. The before-shock image in Fig. 2b shows XRD from the initial material at ambient conditions, with diffraction rings that encode the structure of the initial fcc and bcc phases (red and blue respectively). The diffraction rings marked by the red arrows on the 99 ± 18 GPa image and blue arrows on the 224 ± 30 GPa image denote the compressed peaks plotted in the integrated diffraction traces in Fig. 2 a-b. At 99 ± 18 GPa, we observed XRD peaks of compressed fcc (111) and fcc (200). We observed no other peaks corresponding to another compressed phase, indicating full conversion to the fcc structure at 99 ± 18 GPa. At shock pressures of 210 ± 29 GPa and above, we only observed diffraction peaks from a compressed bcc phase. At shock pressures of 308 ± 39 GPa and above, we observe a broad and diffuse peak typical of a liquid, with no sharp peaks from a crystalline solid; this indicates complete melting. Raw data from all XRD images and profiles are included in Supplementary Fig. S4.

The line profiles shown in Fig. 2a and Supplementary Fig. S5 are obtained by azimuthally integrating the XRD signals, with vertical lines denoting the compressed peaks that convey the phase transformations. The data for each *d*-spacing plotted in Fig. 2b was measured using a peak-fitting algorithm, where the mean value and full-width half-max of each peak corresponds to the d-spacing and uncertainty. The dashed lines in Fig. 2b are calculated assuming isotropic compression along the principle Hugoniot to convert density to a change in *d*-spacing for each lattice plane (supplementary Fig. S6). While the fcc (111) peaks match well with the prediction curves based on the Hugoniot EoS, the fcc (200) peaks are consistently larger than predicted, suggesting possible microstructural dynamics.

The shock pressures are estimated from the results of the line-VISAR measurements and applied laser intensities (See supplementary Information). The Hugoniot relation (fig. 4) of the AM EHEA $AlCoCrFeNi_{2.1}$ was determined by combining the previous Hugoniot data reported by Katagiri *et al*[15] with ambient sound speed data collected in this study. Ambient longitudinal and transverse sound speeds of the AM EHEA were measured on 2 mm-thick samples at the Dynamic Integrated Compression Experimental (DICE) Facility at Sandia National Laboratories, and are shown in Supplementary Table III. The longitudinal sound speed ($V_l$) and transverse sound speed sound speed ($V_s$) were converted to the bulk sound speed ($C_B$) via the following equation: $C_B = \sqrt{{V_l}^2 - \frac{4}{3}{V_s}^2}$. We empirically determined the shock Hugoniot as a linear relationship between the shock velocity ($U_s$) and particle velocity ($u_p$).

**B. Shock MD simulations**

To predict the temperature, we preformed MD simulations to complement our shock experiments. We simulated the fcc phase of $AlCoCrFeNi_{2.1}$ under shock compression to represent a single nanolamella at pressures of 81-420 GPa to match the laser shock experiments. MD simulation results are summarized in Fig. 3 and Table I. The MD simulations describe shorter time scales than what is measured in the ns-timescale shock experiments, however we are able to capture the dynamics of the transition from fcc to bcc to melt in the MD.

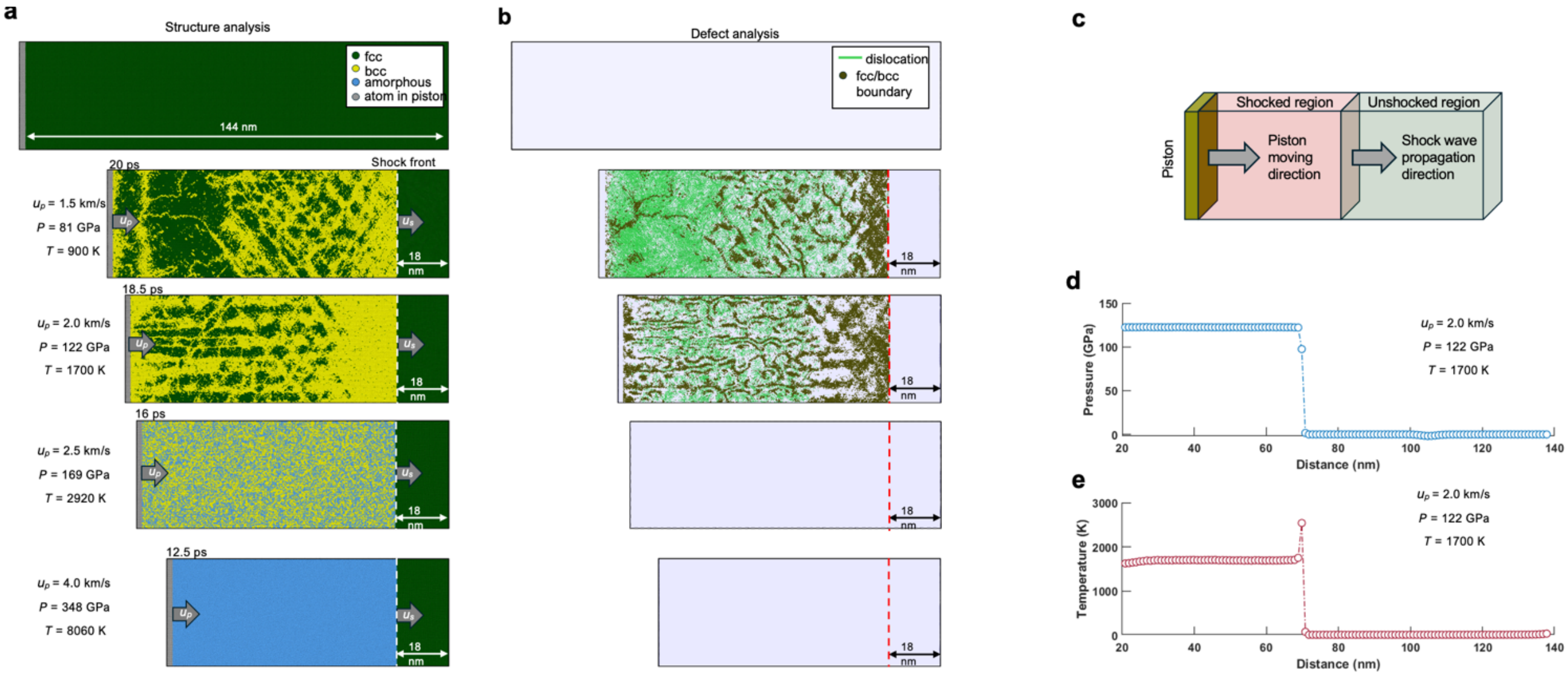

**Figure 3.** (**a**) MD simulation results of shocked AM EHEA after 10 ps. The piston (particle) velocities are 1.5, 2.0, 2.5, and 4.0 km/s from the top to bottom. FCC, bcc, and amorphous structures are colored by green, yellow, and light blue, respectively. (**b**) MD simulation results of shocked AM EHEA after 10 ps. Dislocations are labeled in green. Phase boundaries are labeled in brown (**c**) Schematic of simulation setup with piston, shocked, and unshocked regions denoted by yellow, red, and green respectively. (**d**) Pressure in the material for a particle velocity of 2 km/s. (**e**) Temperature in the material for a particle velocity of 2 km/s.

Table I. Summary of the shock state in the MD simulation results.

| Particle velocity $u_p$ [km/s] | Pressure $P$ [GPa] | Temperature $T$ [K] | Density $\rho$ [g/cm3] |
|---|---|---|---|
| 1.5 | 81 | 899 | 9.36 |
| 2.0 | 122 | 1698 | 9.95 |
| 2.5 | 169 | 2925 | 10.5 |
| 3.0 | 223 | 4616 | 11.0 |
| 3.5 | 282 | 5816 | 11.4 |
| 4.0 | 348 | 8054 | 11.8 |
| 4.5 | 420 | 10990 | 12.2 |

At very early time scales (ps) the simulations suggest that we see the material transform from fcc to bcc followed by a back-transformation from bcc to fcc at lower pressures as demonstrated by the greater population of bcc following the shock front decreasing moving to the left (fig 3a). At higher pressures (>169 GPa in simulations, >200 GPa in experiment), the fcc to bcc transformation is complete, in the MD simulations, we see a coexistence of the bcc and amorphous phase. This amorphization was confirmed using wave optics simulations (fig. S7).

We observe a back transformation from bcc to fcc in the MD simulations that is not seen experimentally. While measurements at the 10-ps timescales relevant to observe the back-transformation was not feasible in our experiment, shock experiments can provide insights into the phase diagram of this material. Our simulation predicted a bcc-dominant structure at 122 GPa, in contrast to the fully fcc structure observed experimentally between 99 ± 18 and 198 ± 24 GPa. Figure 3b depicts the dislocation density and phase boundaries in the fcc phase of our material. No dislocations are found in the bcc phase. Dislocation analysis fails at pressures of 169 GPa due to the amorphization of the material and limited solid regions.

## C. Discussion

Fig. 4 shows our compiled pressure-density equation of state of the nanolamellar AM $AlCoCrFeNi_{2.1}$ from the results in this work. Red, blue, and yellow filled squares are fully fcc, fully bcc, and shock-melted (liquid) phases observed by XRD in our laser shock experiments. We note that the shocked material densities both the MD simulations and the experimental data up to the melt are within the error bars of our Hugoniot measurement.

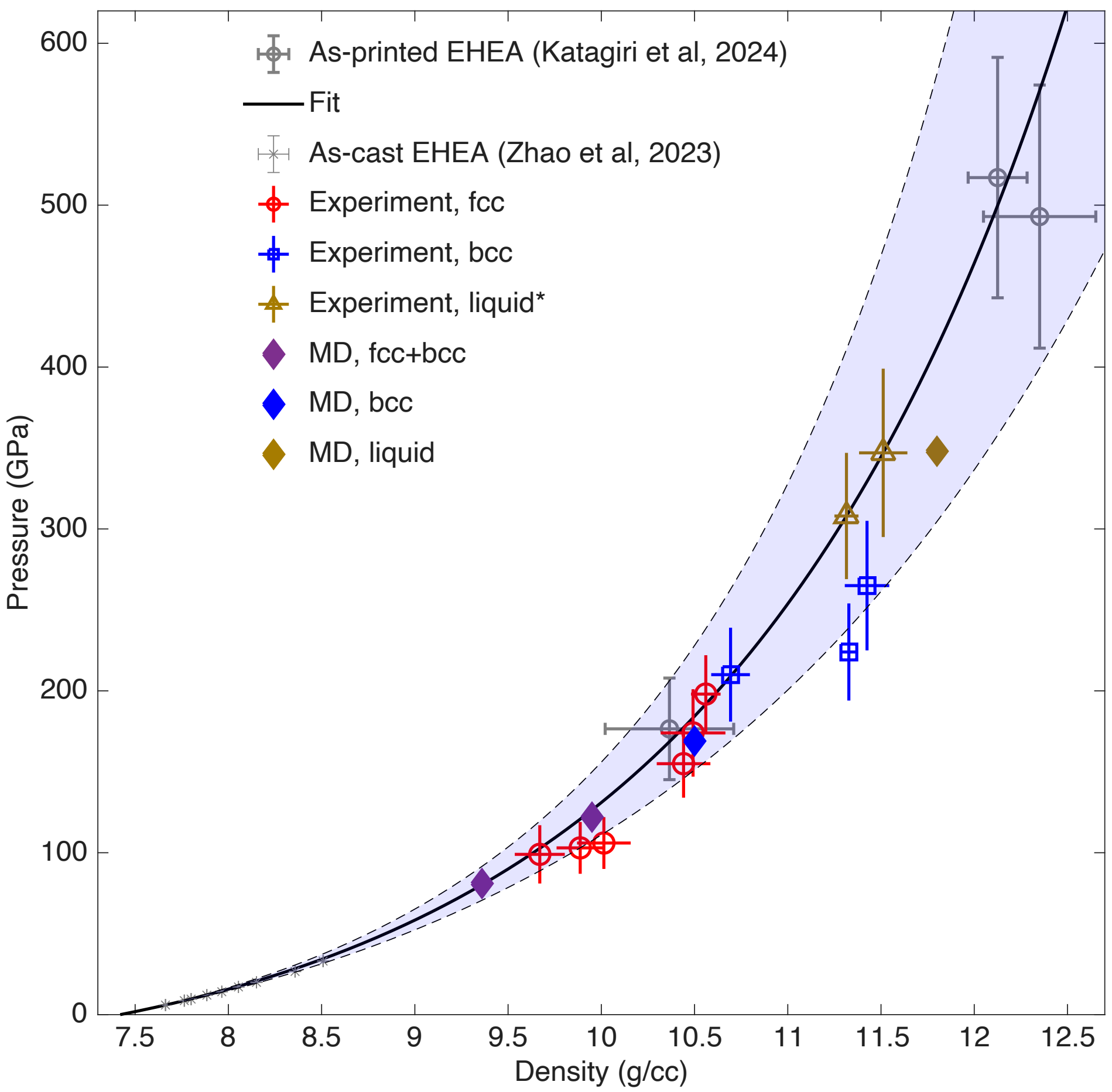

**Figure 4: Pressure-Density diagram of AM-EHEA AlCoCrFeNi2.1.** The black line and shaded region represent the Hugoniot relation determined by fitting results from previous shock experiments on this EHEA together with wave-speed measurements taken at Sandia National Laboratories[15]. Pressures for the experimental points are estimated by using this Hugoniot relation to convert particle velocity measured with VISAR, while densities are determined using the XRD results. For experimental points in the liquid phase, density is also estimated from the Hugoniot relation of Katagiri 2024.

The dynamic compression results show the same dual phase fcc/bcc to fcc only below 100 GPa. The fully fcc phase is observed up to 198 ± 24 GPa which is consistent with previous DAC measurements at room temperature[15], suggesting a large range of thermal stability of the fully-fcc phase at this pressure.

In our experiments, we observe a transition to bcc at to 198 ± 24 GPa. This transition to bcc only at pressures around 200 GPa is not observed in the statically compressed EHEA-1 data[15], implying inertial or thermodynamical effects governing the transition to bcc. Our observation that the entropically-preferred bcc phase is observed just before melting is observed shock data, as seen in other metals[21–23]. This is thought to be due to bcc being a less close-packed and thermodynamically favored phase at high pressures, not a kinetic effect. At the high pressures relevant to this phase transition, it is expected that dislocation mediated flow is not the key driver of plasticity. Rather at these high pressures, strain relaxation is expected due to phase transformations[21].

Given the coupled nature of temperature and strain during a shock compression experiment, we turn to MD simulations and previous experiments to elucidate the effects of the competing thermal and mechanical processes during shock

As depicted in Fig. 1d-e, the MD simulations show a spike in temperature behind the shock front and a consistent rise in temperature in the shocked material. A commensurate spike in pressure behind the shock front is not observed. Fig. S7 describes the interplay between heating and strain during shock compression..

As shown in Fig, 1b, our MD simulations demonstrate a high dislocation density in the fcc phase at lower pressures, with the density of dislocations decreasing as the pressure increases, increasing the phase boundaries present. These MD simulations give potential insights into the strain release mechanisms between the fcc and bcc phases of this material. This behavior has been previously studied in analogous alloy -- the cast EHEA $AlCoCrFeNi_{2.1}$ – by Shen et al. This study used *in-situ* XRD and *ex-situ* electron backscatter diffraction (EBSD) to study the deformation behavior of the cast alloy under tensile testing. This material exhibited stress partitioning with fcc showing a higher dislocation density at the onset of plasticity with plastic deformation starting in only fcc ordered regions until the effective stress transfer to bcc/B2 occurs. As strain is increased, the dislocation density in both the fcc and bcc/B2 regions increases, with the bcc/B2 regions preventing the free deformation of the fcc phased regions. This study demonstrated a larger lattice strain in the bcc/B2 regions than in the fcc phase due to partitioned stress behavior of each phase. As seen in figure 1d, the observed *d*-spacings of the fcc (111) peaks match well with the prediction curves based on the Hugoniot equation-of-state, while the fcc (200) peaks are consistently larger than the prediction. Differences in the compressibility between different diffraction planes of the same phase has previously been seen in shocked metals and interpreted as the effect of stacking fault formation[21,24]. This observation suggests a potential defect mediated plasticity mechanism in the fcc phase. Together with our MD simulations, we see that the fcc phased material releases strain through the nucleation of dislocations. We observe no dislocations in the simulations of our bcc phase, indicating a limited ability of this phase to accommodate plastic deformation through defect formation (Fig. 1b).

In our MD simulations, we further can elucidate the material response to compression by computing the local instantaneous kinetic temperature. This allows us to resolve transient heating during uniaxial shock compression and to quantify its temporal correlation with the onset of plasticity. We observe the onset of amorphization alongside the complete transition from fcc to bcc, indicating potential thermal effects in this transition (Fig. 1a, S7, S8). Our simulations, shown in Figure S7, demonstrate a temperature spike behind the shock front with a lower temperature plateau further behind the shock. These simulations predict that strain is inhomogenously distributed in the sample, with strain segregation around phase boundaries. Due to the coupled nature of temperature and strain during a shock wave, elucidating the temperature versus strain dependencies of the observed phase transitions experimentally requires the use of heated static compression where thermal and pressure effects can be isolated from strain.

The time scale difference between shock deformation (nanoseconds) and MD simulations (picoseconds) would also result in differences in the deformation system of the material. Although the simulated pressure-density relationship agrees with the experimental results, the shock-induced phase transitions predicted in our MD simulations differ from our XRD results. The MD simulations predicted a mixture of bcc and fcc at a shock pressure of 81 GPa, while our experiments do not observe this. Additionally, our simulation predicted a bcc-dominant structure at 122 GPa, in contrast to the fully fcc structure observed experimentally between 99 ± 18 and 198 ± 24 GPa. At 161 GPa, simulations predict the coexistence of bcc and amorphous regions. The transition to liquid is complete by a shock pressure of 348 GPa in agreement with the experimental data. Since the potential used for the simulations was not initially designed to study high-pressure high-temperature conditions, the experimental data can be used as a benchmark for further improvements to this potential.

Classical molecular dynamics necessarily omits some physical interactions for the sake of computational efficiency. Notably, it cannot represent electronic excitation and electron-mediated heat transport during shock compression in order to remain tractable at the scale of millions of atoms. The simulations also employ finite cutoff radii, a semi-empirical interatomic potential, and periodic boundary conditions transverse to the shock direction to approximate bulk behavior. These choices truncate long-range pairwise interactions and confine lateral fluctuations via periodicity while relying on a fitted potential of limited transferability, which can bias the equation of state (altering shock speed and heating), and defect/microstructure evolution. These limitations combined with inherently short MD timescales and the high strain rates required for feasible simulations complicates a direct, quantitative comparison between our MD results and bulk shock responses measured by XRD. Nevertheless, MD resolves individual atomic trajectories on femtosecond-to-picosecond timescales, providing mechanistic insight into the unshocked-to-shocked transition.

The difference in phase between MD simulation and dynamic experiments is expected since the potential for this material is not well validated at high strain rates due to the lack of experimental data. By providing high strain rate phase data, we can improve the predictive ability of these simulations. Since these EHEAs are designed for extremes, future work using this data can improve the predictive ability of these simulations and design better materials for applications at extremes.

As our shock measurements did not include the range of d-spacings characteristic of the B2 phase, we cannot comment on inertial effects' change of the phase ordering. Our XRD for the shock experiments is not sensitive to detect the difference between B2 and bcc because of its limited $2\theta$ detection range. Thus, the phase we identify as bcc in the shock experiments can possibly be B2 phase or mixture of both bcc and B2. Further targeted experiments are required to elucidate the formation of intermetallic phases under shock compression.

Future work needs to be done to characterize the effects of inertia on phase. Shock waves can access pressures difficult to reach in a quasi-static experiment; however, several previous studies using shock to determine phase have demonstrated characterizable inertial effects[25]. Since this is the first high pressure phase diagram of AM $AlCoCrFeNi_{2.1}$, this study enables further static studies to characterize the impact of kinetics on the high-pressure response of this material.

## III. Conclusion

We conducted dynamic compression experiments to measure structural transitions of AM $AlCoCrFeNi_{2.1}$ under high pressures. The XRD results of the dynamically compressed AM EHEA captured phase transitions along its Hugoniot, starting from the initial fcc/bcc eutectic structure. Under compression, the structure first transitions to a fully fcc phase, followed by a fully bcc phase at higher pressures, before reaching melting at even higher pressures. Our results provide the first phase diagram of the AM nanolamellar AM $AlCoCrFeNi_{2.1}$ under a wide range of pressures for possible operating conditions. Understanding the high-pressure behavior of this material lays the groundwork for industrial applications at extremes. From our results, microstructural studies can now explore how the phase transformations we have characterized can result in unique strengthening effects.

## Methods

### Material synthesis

The $AlCoCrFeNi_{2.1}$ EHEA samples used in this work were additively manufactured using a commercial EOS M290 laser powder bed fusion system with an ytterbium-fiber laser with a focal diameter of 100 μm and a maximum power of 400 W. Gas(argon)-atomized $AlCoCrFeNi_{2.1}$ powders (Vilory Advanced Materials Technology) with particle sizes ranging from 15 μm to 53 μm with a mean value of 35 μm were used as the starting material. The printing was done in an argon atmosphere with oxygen concentration of < 1,000 ppm. A 4140-alloy steel plate was used as the printing substrate, and it was preheated to 80 °C to mitigate the build-up of thermal residual stresses. The printing was conducted using laser power of 350 W, scan speed of 1000 mm/min, layer thickness of 0.04 mm, and hatch spacing of 0.08 mm. A bi-directional scan pattern was used with a 90º rotation between each layer. The density of the as-printed EHEA was determined to be 7.42 g/cm$^3$ using gas pycnometry. See Ren *et al*[12] for more detail about fabrication method and characterization of the printed material.

### Dynamic compression measurements using laser-shock technique

For the laser-shock experiment, the printed samples were cut into a 3 mm diameter disk using a disk punch. Then they were thinned to ~20 μm by hand-polishing both surfaces. Each sample was attached onto a polypropylene ablator (see Supplementary Table I for their thickness). Then we used a laser scanning microscope (VK-9700, Keyence) to scan the EHEA step height from the ablator surface to characterize the EHEA thickness and its uniformity. The polished surfaces were cleaned using acetone and anhydrous ethanol to remove oil stains and then blow-dried after cleaning.

The *in-situ* X-ray diffraction (XRD) measurements of laser-shocked AM EHEA $AlCoCrFeNi_{2.1}$ were performed at SACLA[17], the X-ray Free Electron laser (XFEL) in Japan. A high-energy nanosecond optical laser was spatially and temporally synchronized to the XFEL at BL3-EH5[18]. The angle between the sample normal and XFEL irradiation direction ($\chi$ in Fig. 1a) was set at 45 deg for most of the shots except the three highest-pressure shots which used $\chi$ = 18 deg instead. Three different photon energies, 10.00 keV ($\lambda$ = 0.1240 nm), 11.12 keV ($\lambda$ = 0.1115 nm), and 12.00 keV ($\lambda$ = 0.1033 nm), all with full width at half maximum of ~0.03 keV, were used in the self-amplified spontaneous emission (SASE) mode. The pulse duration of the XFEL was ~10 fs and the pulse energy was >300 μJ/pulse. The X-ray beam diameter was between 20-50 μm and thus it collects diffractions from numbers of the nano-lamellae. The X-rays diffracted by the sample were recorded by a two-dimensional (2D) flat panel X-ray detector. The detector positions were calibrated using diffraction patterns from a $CeO_2$ powder taken at ambient conditions. The collected XRD images were analyzed using IPAnalyzer[26], PDIndexer [27], and ReciPro[28].

The high-energy optical laser delivered a nanosecond pulsed beam focused onto the ablator to drive a strong shock wave. As the shock wave propagates in the ablator then the AM EHEA, it compresses the materials to high-pressure and high-temperature with a high strain-rate of $\geq 10^8$ $s^{-1}$. The drive laser irradiation side of the ablator had a black ink layer to block the pre-heating of the sample by the laser radiation. This makes the temperature rise in the sample to be purely the effect of the shock deformation. The laser pulse was shaped to form a quasi-square pulse with 5 ns duration with a ≤100 ps rise time. To apply different peak pressures in the EHEA sample, we used the attenuator upstream of the laser path to vary the on-target laser energy between 1 and 18 J. At SACLA, the spatial laser pattern on the shock target was smoothed using diffractive optical elements (DOE). We used two DOEs with different focal spot sizes, 260 and 180 μm in diameter, which also contributed to varying the peak shock pressures in the EHEA.

Simultaneous to the XRD measurements, we used line-VISAR[19,20] to estimate the shock pressure in the EHEA sample via impedance matching. A single crystal [100] lithium fluoride (LiF) or z-cut quartz window was used for some of the shots to improve the accuracy of the shock-state characterization by VISAR. When a lithium fluoride window was used, the line-VISAR tracked the apparent velocity ($u_{app}$) of the interface between the AM EHEA and LiF crystal, through the compressed LiF. The particle velocity of AM EHEA ($u_p$) is obtained from measured $u_{app}$ using the known refractive index correction factors from ref[29]. When a quartz window was used, the line-VISAR time-resolved the shock velocity ($u_s$) in the quartz. The $u_p$ of AM EHEA is then determined using the AM EHEA $AlCoCrFeNi_{2.1}$ Hugoniot and quartz Hugoniot in[30]. The AM EHEA $AlCoCrFeNi_{2.1}$ Hugoniot was reported in[15], but we updated the Hugoniot using ambient sound speed data we collected in this work (see Supplementary Information). The Hugoniot of the AM EHEA $AlCoCrFeNi_{2.1}$ was assumed to follow the standard linear relationship between the shock and particle velocity: $U\text{s} = 4.759(\pm 0.085) + 1.646(\pm 0.149)\, u_p$. We assumed the release path of the EHEA to be identical to the reflected Hugoniot to perform impedance mismatching[31] between the EHEA and quartz since no release data is available for the AM EHEA. For the shots without LiF or quartz, the line-VISAR tracked the free-surface motions of the AM EHEA sample ($u_{fs}$). The $u_p$ in this case was again assumed the release was approximated by the reflected Hugoniot: $u_{fs} = 2u_p$[32,33]. In some relatively high-pressure shots without a window material, a significant drop in reflectivity upon shock breakout was observed and we were unable to accurately measure the free surface velocity. In these cases, and for a few low-pressure shots with low line-VISAR quality, we estimated the peak shock pressure in the AM EHEA from the measured laser intensity, using the relationship between the laser intensity delivered to the target and peak shock pressure determined by VISAR for other shots (Supplementary Fig. 2). The uncertainties for the peak shock pressures reported in this work include the uncertainties of the line-VISAR measurements, as well as the uncertainties of the Hugoniot data of LiF, quartz, and the AM EHEA.

**MD simulations of the shocked EHEA**

We conducted large-scale MD simulations to explore the dynamic shock compression behavior in $AlCoCrFeNi_{2.1}$. The simulations were carried out using the open-source Large-scale Atomic/Molecular Massively Parallel Simulator (LAMMPS) code[34]. For the simulation cell, we constructed a fcc $AlCoCrFeNi_{2.1}$ single crystal with an approximate size of 37 x 146 x 37 nm$^3$, containing 16 million atoms. We chose the single crystal, instead of the eutectic polycrystal used in the experiments, because the cell size was close to the size of the AM EHEA nanolamellae. The x, y, and z axis were oriented along [100], [010], and [001] directions of the fcc crystal, respectively. Free boundary conditions were applied to y axis (shock loading direction) and periodic boundary conditions were applied to x and z axis. A many-body embedded atom method (EAM) potential[35] was employed for the $AlCoCrFeNi_{2.1}$ single crystal. To reduce the residual stress, the single crystal was equilibrated at a temperature of 10 K for 5 ps before loading with a time step of 1 fs in an isobaric-isothermal (NPT) ensemble. The shock process was performed using the non-equilibrium molecular dynamics (NEMD) method. A slab of atoms, without any interaction force between them, on the left-hand side of the single crystal having dimensions of 37 x 2 x 37 nm$^3$ was selected as the piston. A planer shock compression wave was generated by moving the piston at a constant velocity of $V_P$ ranging from 1.0 - 4.5 km/s along y axis for 10 ps.The physical properties during shock compression were determined using binning. We divided the single crystal into 1 nm wide intervals or bins along the y axis.

**Data availability**

All data that support the findings of this study are available from the corresponding authors on reasonable request.

**Acknowledgements**

This material is based upon work supported by the Department of Energy—National Nuclear Security Administration Center of Excellence CAMCSE under Award No. DE-NA0004154. The laser-shock experiments were performed at BL3 of SACLA with the approval of the Japan Synchrotron Radiation Research Institute (proposal nos. 2024A8007, 2023A8016, and 2022B8026). The part of the results was obtained using a high-power nanosecond laser deployed at SACLA by Institute of Laser Engineering of Osaka University with the corporation of Hamamatsu Photonics. The installation of Diffractive Optical Elements (DOE) to improve the smoothness of the drive laser pattern was supported by the SACLA Basic Development Program. Sandia National Laboratories is a multimission laboratory managed and operated by National Technology \& Engineering Solutions of Sandia, LLC, a wholly owned subsidiary of Honeywell International Inc., for the U.S. Department of Energy's National Nuclear Security Administration under contract DE-NA0003525. This paper describes objective technical results and analysis. Any subjective views or opinions that might be expressed in the paper do not necessarily represent the views of the U.S. Department of Energy, the National Transportation Safety Board, or the United States Government. We thank Andrew Pope from University of Alabama Birmingham for his contributions and discussions.

**Author contributions**

K.K., Y.V., W.C., and L.D.-M. conceived this study. J.R., W.Y., W.C. additively manufactured the materials. K.K. and A.H. fabricated the shock targets. K.K., A.H., S.P., T.R., S.J.I., L.M., N.O., A.A., R.K., H.N., Y.N., M.O., Y.S., S.T., T.O., Y.U., Y.I., K.M., K.S., T.T., M.Y., T.Y., P.E.S., Y.K.V., and L.D.-M. conducted the experiments. H.C., W.L., D.L. and P.C. performed and analyzed the MD simulations. K.K. and S.E.P. analyzed the experimental data and wrote the paper. All authors reviewed and approved the manuscript.

**Competing interests**

The authors declare no competing interests.

**Correspondence and requests for materials**

should be addressed to Kento Katagiri or Leora Dresselhaus-Marais.

**Supplementary Information**

**Phase transitions of eutectic high entropy alloy $AlCoCrFeNi_{2.1}$ under shock compression**

Sophie Parsons[1,2,3], Kento Katagiri[1,2,3*], Hangman Chen[4], Andrew D. Pope[5], Anirudh Hari[1,2,3], Tharun Reddy[1,2,3], Sara J. Irvine[2,3,6], Laura Madril[1,2,3], Dorian Luccioni[1,2,3], Jie Ren[7], Wuxian Yang[7], Norimasa Ozaki[8,9], Alexis Amouretti[8], Ryosuke Kodama[8,9], Hirotaka Nakamura[8], Yusuke Nakanishi[8], Masato Ota[10], Yusuke Seto[11], Sota Takagi[12], Takuo Okuchi[13], Yuhei Umeda[13], Yuichi Inubushi[14,15], Kohei Miyanishi[15], Keiichi Sueda[15], Tadashi Togashi[14,15], Makina Yabashi[14,15], Toshinori Yabuuchi[14,15], Wanghui Li[16], Paul E. Specht[17], Penghui Cao[4], Wen Chen[7], Yogesh K. Vohra[5], Leora E. Dresselhaus-Marais[1,2,3,*]

[1]Department of Materials Science and Engineering, Stanford University, California, USA.

[2]SLAC National Accelerator Laboratory, California, USA.

[3]PULSE Institute, Stanford University, California, USA.

[4]Department of Mechanical and Aerospace Engineering, University of California, Irvine, California, USA.

[5]Department of Physics, University of Alabama at Birmingham, Alabama, USA.

[6]Department of Applied Physics, Stanford University, California, USA.

[7]Department of Mechanical and Industrial Engineering, University of Massachusetts, Massachusetts, USA.

[8]Graduate School of Engineering, Osaka University, Osaka, Japan.

[9]Institute of Laser Engineering, Osaka University, Osaka, Japan.

[10]National Institute for Fusion Science, Gifu, Japan.

[11]Graduate School of Science, Osaka Metropolitan University, Osaka, Japan.

[12]Earth and Planets Laboratory, Carnegie Institution for Science, Washington DC, USA.

[13]Institute for Integrated Radiation and Nuclear Science, Kyoto University, Osaka, Japan.

[14]Japan Synchrotron Radiation Research Institute, Hyogo, Japan.

[15]RIKEN SPring-8 Center, Hyogo, Japan.

[16]Institute of High Performance Computing (IHPC), Agency for Science, Technology and Research (A*STAR), Singapore 138632, Republic of Singapore

[17]Sandia National Laboratories, Albuquerque, New Mexico, USA.

[18]These authors contributed equally: Kento Katagiri and Sophie Parsons

*Corresponding authors. Email: kentok@stanford.edu & leoradm@stanford.edu

## I. Line-VISAR measurements for shock pressure estimation

Summary of the shock experiment results are shown in Supplementary Tables I and II. As described in the method section, the shock pressures in this work were estimated from the velocities measured by the line-VISAR or the on-target laser intensity, as describe in the Method section. A typical line-VISAR image obtained in this work is shown in Supplementary Fig. 1 along with the velocity profiles. The line-VISAR image shown in Supplementary Fig. 1a shows the phase shifts due to the shock wave breaking out from the free-surface of the AM EHEA sample, meaning no window was used for the shot. A standard push-pull line-VISAR with two different velocity per fringe (VPF) constants was used and Supplementary Fig. 1a shows the record for only one VPF. Supplementary Fig. 1b shows the velocity traces determined from both images generated by the line-VISAR images. The line-VISAR image shown in Supplementary Fig. 1a indicates the shock wave is spatially flat for > 200 μm and this is much wider than the XFEL spot size (≤ 30 μm), meaning the X-ray is probing the compressed volume of the sample with no effect from the edge waves. The x-ray measurements probe the center of the compressed region, whereas the center of the VISAR measurement is intentionally offset from the center of the compressed region to capture the shock-releasing edge as seen in the left-hand side of the VISAR image (Supplementary Fig. 1a).

As indicated in Supplementary Table I, shock pressures for some of the shots were estimated from the measured laser intensity. We first obtained the relationship between the laser intensity and the peak shock pressure for some shots at different pressures using line-VISAR, and the results are shown in Supplementary Fig. 2. To develop this relationship between laser intensity and shock pressure in the AM EHEA, we performed additional shots collecting only line-VISAR data without the XFEL.

Due to the scarcity of experimental Hugoniot data of the AM EHEA, we used our experimentally determined bulk sound speed to constrain the intercept of the linear fit, and the three Hugoniot points of Katagiri *et al*[15] to determine the slope of the line. The fitting was weighted according to the uncertainties of the individual data points, and the uncertainty of the fit accounts for the uncertainties in the data points. See Supplementary Fig. 3 for the obtained Hugoniot equation-of state.

This approach differs from the method that Katagiri *et al* used for their linear fit. Katagiri *et al* fit their data of AM $AlCoCrFeNi_{2.1}$ together with data of as-cast $AlCoCrFeNi_{2.1}$ from Zhao *et al*[36], without accounting for the uncertainties of the data points in the fit. The EHEA used in Zhao's experiment was made by casting method and its lower cooling rate resulted in micron-sized lamellae, in contrast to the nanolamellae formed by additive manufacturing. Katagiri *et al* had to neglect the difference of the microstructure to use Zhao's data as they did not have the ambient sound speed data to constrain the intercept. We propose that the method we use in this paper is more robust since we constrain the intercept, include the uncertainties of the data points, and only use data taken on the AM EHEA.

The line VISAR shock Hugoniot measurements were used to estimate the density of the liquid phase EHEA-1. This measurement was verified using Liquid Diffract[CITE].

## II. XRD patterns obtained in the laser shock experiments

All XRD patterns collected but not presented in the article are shown in Supplementary Fig. 4. The shot numbers and estimated shock pressures are noted on each image. Note that three different photon energies of 10.0, 11.1, and 12.0 keV were used in this work, as indicated in Supplementary Table I, and this resulted in the XRD peaks from the same crystal plane to appear at different scattering angle ($2\theta$). Except for the two highest pressure shots of 308 ± 39 GPa and 347 ± 52 GPa, the X-ray probe was done while the shock wavefront was within the AM EHEA sample, meaning the X-ray is collecting diffractions from both the compressed and uncompressed AM EHEA. For the two highest pressure shots, however, we set the X-ray probe timing very close to the timing of the shock-wave breaking out from the rear surface of the AM EHEA, to capture the weak diffraction signals from the melted AM EHEA. The weak spots seen in the diffraction patterns are from the single crystal windows.

The XRD at 265 ± 40 GPa observed a strong and spotty ring from compressed bcc (110) (See Supplementary Fig. 6a also). While this usually indicates formation of a large bcc grain, we did not observe a similar feature in and other XRD patterns. The cause of this is still being investigated.

## III. Bcc phase observed under shock compression

As described in the main paper, we interpret the phase observed at shock pressures of 210 ± 29, 224 ± 30 GPa, and 265 ± 40 GPa to be the bcc phase. Although only the bcc (110) peak was clear in the XRD line profiles (See Fig. 2a for the 224 ± 30 GPa shot and Supplementary Fig. 5 for the other two shots), careful background correction of the data allowed us to confirm that there are two additional peaks of bcc (200) and bcc (211) for 210 ± 29 and 224 ± 30 GPa data (Supplementary Fig. 6b&c). For the 265 ± 40 GPa shot, however, we could not confirm bcc (200) peak even after the background subtraction and the bcc (211) peak was outside of the observation range, as a low photon energy of 10.0 keV was used for the shot. The absence of the bcc (200) peak for 265 ± 40 GPa shot could be explained by the formation of large bcc crystals, which is supported by the intense spotty bcc (110) ring observed.

**Supplementary Table I. Summary of the shock experiment results.**

| Shot No. | Ablator thickness [μm] | Window material | Photon energy [keV] | Laser intensity [TW/cm$^2$] | VPFs [km/s /fringe] | Pressure estimated from laser intensity |
|---|---|---|---|---|---|---|
| 1278171 | 30 | LiF | 12.00 | 1.55 | 3.922/ 5.382 | |
| 1188844 | 50 | None* | 11.12 | 1.99 | 5.375/ 8.597 | 106 ± 16 |
| 1188852 | 50 | None | 11.12 | 2.32 | 5.375/ 8.597 | |
| 1278420 | 30 | LiF | 12.00 | 2.40 | 3.922/ 5.382 | |
| 1410598 | 50 | LiF* | 12.00 | 4.20 | 5.419/ 2.894 | 174 ± 27 |
| 1188865 | 50 | None | 11.12 | 4.40 | 5.375/ 8.597 | |
| 1278176 | 30 | None | 12.00 | 3.60 | 3.922/ 5.382 | |
| 1278424 | 30 | LiF | 12.00 | 4.72 | 3.922/ 5.382 | |
| 1279020 | 30 | None* | 10.00 | 7.85 | 3.922/ 5.382 | 265 ± 40 |
| 1279030 | 30 | Quartz | 10.00 | 11.07 | 3.922/ 5.382 | |
| 1279038 | 30 | Quartz* | 10.00 | 11.72 | 3.922/ 5.382 | 347 ± 52 |

(*) Laser intensity was used to estimate the shock pressure as we could not determine velocity from the line-VISAR for some of the shots. This is due to low reflectivity from the moving surface or ghost fringes from the window rear surface.

**Supplementary Table II. Summary of the line-VISAR measurements and impedance mismatching results.**

| Shot No. | Free surface velocity, $U_{fs}$ [km/s] | LiF apparent velocity, $U_{app}$ [km/s] | quartz shock velocity [km/s] | EHEA $u_p$ [km/s] | EHEA $U_s$ [km/s] | EHEA $P$ [GPa] | EHEA $\rho$ [g/cm$^3$] |
|---|---|---|---|---|---|---|---|
| 1278171 | | 3.11 ± 0.24 | | 1.75 ± 0.23 | 7.64 ± 0.72 | 99 ± 18 | 9.63 ± 0.46 |
| 1188852 | 3.61 ± 0.42 | | | 1.81 ± 0.23 | 7.73 ± 0.73 | 104 ± 16 | 9.68 ± 0.47 |
| 1278420 | | 4.23 ± 0.21 | | 2.39 ± 0.24 | 8.70 ± 0.84 | 155 ± 21 | 10.2 ± 0.5 |
| 1188865 | 5.66 ± 0.33 | | | 2.83 ± 0.22 | 9.42 ± 0.87 | 198 ± 24 | 10.6 ± 0.6 |
| 1278176 | 5.90 ± 0.45 | | | 2.95 ± 0.27 | 9.62 ± 0.97 | 210 ± 29 | 10.7 ± 0.6 |
| 1278424 | | 5.41 ± 0.22 | | 3.07 ± 0.29 | 9.82 ± 1.02 | 224 ± 30 | 10.8 ± 0.7 |
| 1279030 | | | 10.65 ± 0.41 | 3.78 ± 0.33 | 10.98 ± 1.19 | 308 ± 39 | 11.3 ± 0.8 |

**Supplementary Table III. Ambient sound speed measurements of the AM EHEA $AlCoCrFeNi_{2.1}$**

| Sample | Longitudinal sound speed [km/s] | | Transverse sound speed (shear) [km/s] | |
|---|---|---|---|---|
| | Measurement 1 | Measurement 2 | Measurement 1 | Measurement 2 |
| Sample 1 | 5.828 | 5.821 | 2.804 | 2.811 |
| Sample 2 | 5.816 | 5.814 | 2.806 | 2.801 |
| Sample 3 | 5.852 | 5.831 | 2.955 | 2.941 |
| Sample 4 | 5.774 | 5.779 | 2.901 | 2.902 |
| Sample 5 | 5.802 | 5.802 | 2.861 | 2.870 |
| Average | 5.812 ± 0.024 | | 2.865 ± 0.058 | |
| Bulk Sound Speed [km/s] | 4.759 ± 0.085 | | | |

**Supplementary Table IV. Phase fractions and lattice parameters of ambient and shocked data**

| Shot No. | Ambient phase fraction fcc/bcc | Shocked fcc or bcc | a fcc (110) [Å] | a bcc (110) [Å] | a fcc (200) [Å] | a bcc (200) [Å] | a fcc (220) [Å] | a bcc (211) [Å] |
|---|---|---|---|---|---|---|---|---|
| Ambient | 0.402 ± 0.109 | - | 3.592 | 2.868 | 2.541 | 2.031 | 2.541 | 2.341 |
| 1278171 | | fcc | 3.293 | | 2.352 | | 2.335 | |
| 1188844 | | fcc | 3.255 | | 2.348 | | 2.319 | |
| 1188852 | | fcc | 3.268 | | 2.337 | | 2.325 | |
| 1278420 | | fcc | 3.209 | | 2.301 | | 2.277 | |
| 1410598 | | fcc | 3.204 | | 2.300 | | | |
| 1188865 | | fcc | 3.197 | | 2.273 | | 2.260 | |
| 1278176 | | bcc | | 2.528 | | 1.775 | | 2.037 |
| 1278424 | | bcc | | 2.479 | | 1.743 | | 2.024 |
| 1279020 | | bcc | | 2.471 | | | | |

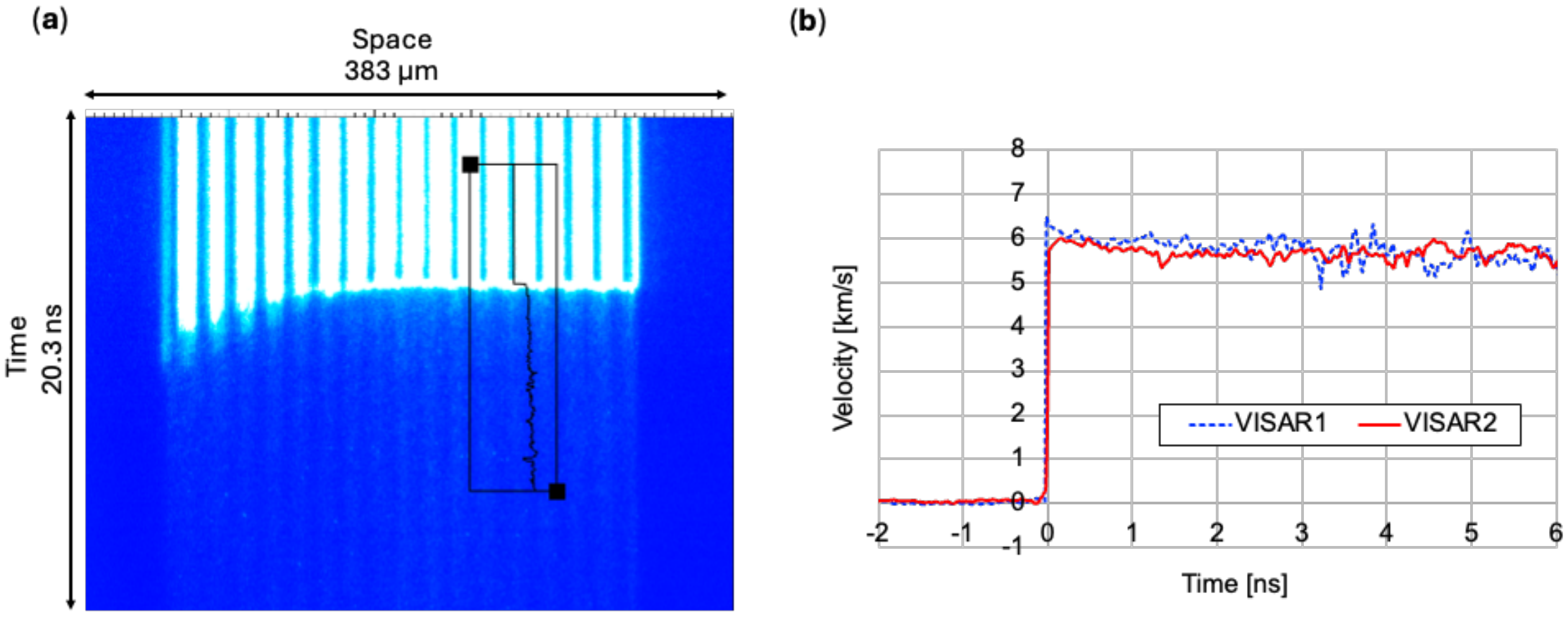

**Supplementary Fig. 1. Typical line-VISAR results obtained for the shock experiments at SACLA.** (**a**) Image of line-VISAR1 for shot number 1278176. (**b**) Velocity profiles of both line-VISAR arms for the same shot.

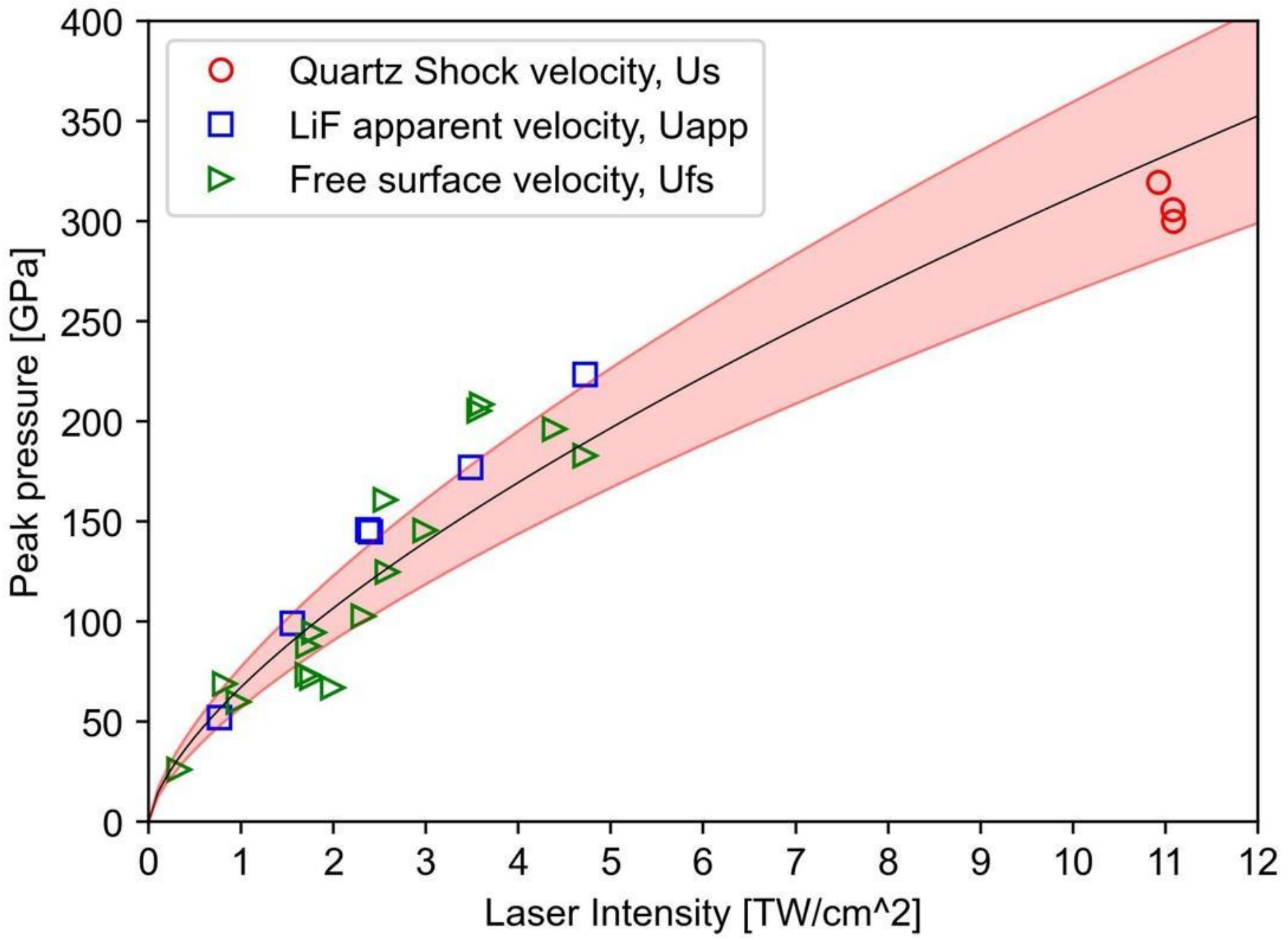

**Supplementary Fig. 2. Laser-intensity versus pressure (stress) relationship for SACLA.** The y-axis is the peak pressure in the AM EHEA obtained from different types of velocities measured by VISAR. The red circles were obtained by measuring the shock velocity ($U_s$) in the quartz window attached to the rear surface (VISAR irradiation side) of the AM EHEA sample. Blue squares represent shots measured the apparent velocity ($u_{app}$) from the interface between the AM EHEA and LiF window. Green plots were obtained by measuring the free surface velocity ($u_{fs}$) of the AM EHEA, meaning no window material was on the AM EHEA sample. Some of the plots are obtained for calibration shots (*i.e.*, without X-ray measurements). The black curve is the fitting curve determined under the constraint $y = ax^{\frac{3}{2}}$. The red shaded area shows the standard deviation of the fit. These fit and standard deviation are used to estimate shock pressures and their errors for the laser shock experiments with poor line-VISAR return.

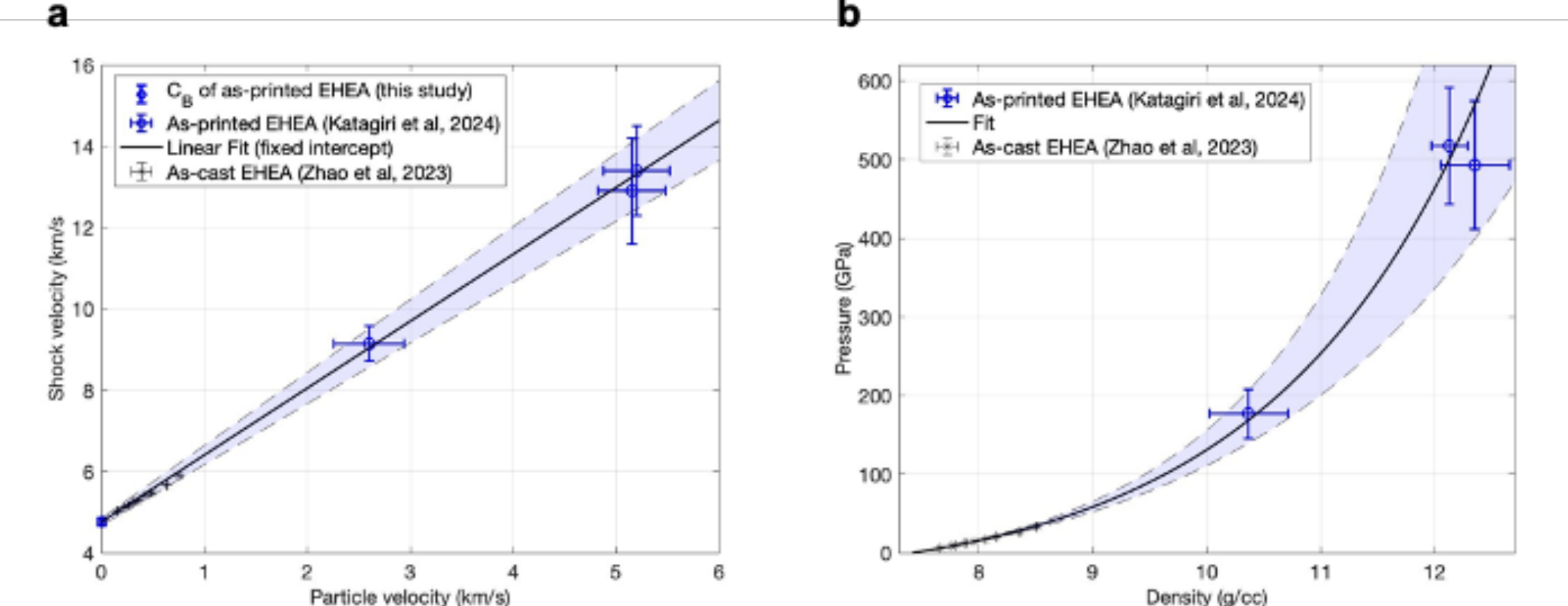

**Supplementary Fig. 3. Hugoniot equation-of-state of AM EHEA $AlCoCrFeNi_{2.1}$.** The blue circles are the Hugoniot data obtained from laser-shock experiments [12] of the AM EHEA with nanolamellar structure. The black crosses are the Hugoniot data measured by plate-impact experiments [29] of as-cast EHEA with the same composition. The conventional casting process results in a micro-lamellar structure rather than the finer nanolamellae found in the AM EHEA. (**a**) The linear $U_S$-$u_p$ Hugoniot relation of AM $AlCoCrFeNi_{2.1}$, fit to the data of Katagiri *et al*. [12] and sound speed data presented in this study. The shaded blue region indicates uncertainty of one standard deviation above and below the line. (**b**) The same Hugoniot relation presented in pressure-density space (identical to Fig. 3c).

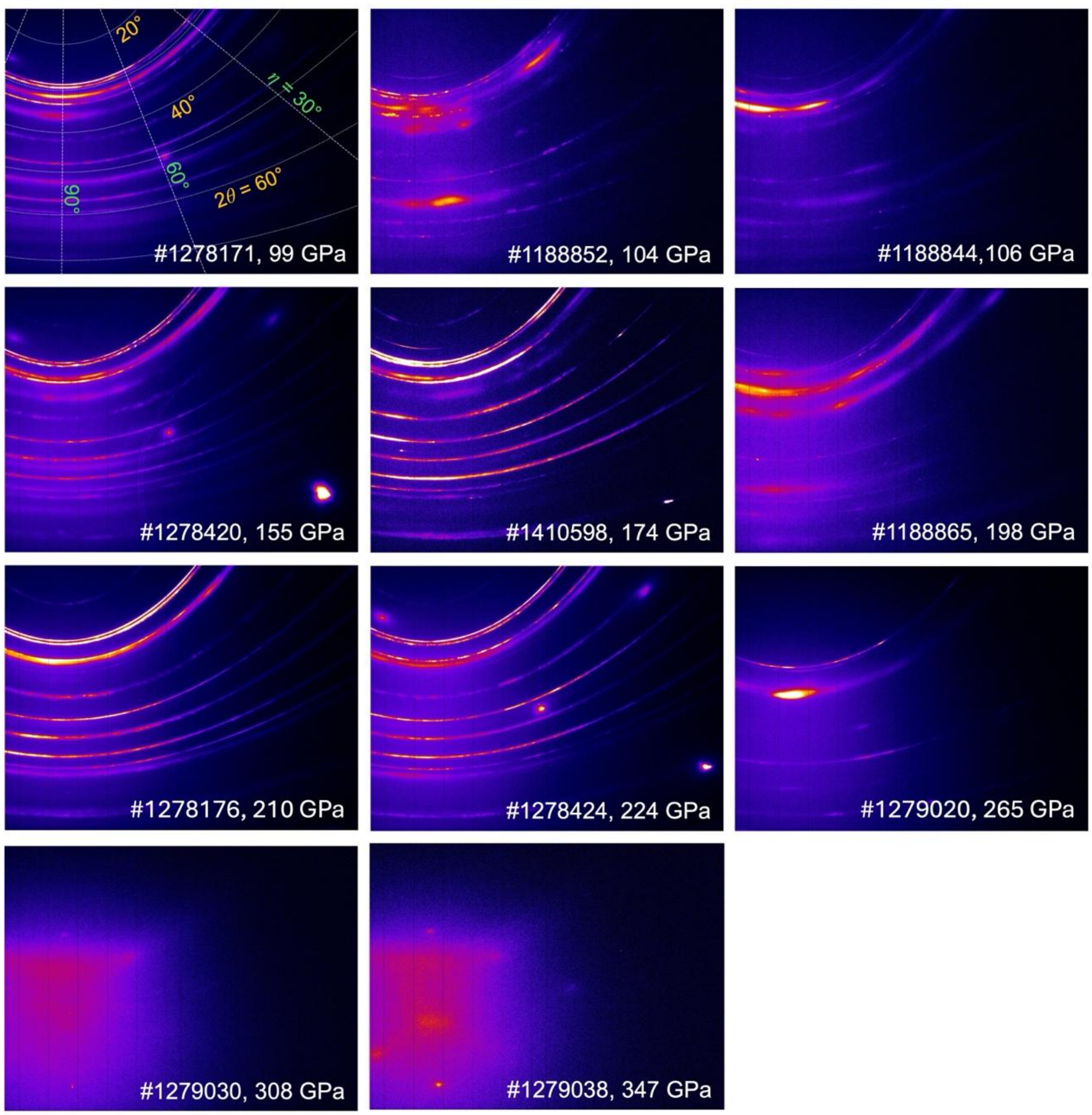

**Supplementary Fig. 4. Summary of the XRD images for the shock experiments.** The shot number and the estimated peak pressure in the AM EHEA are denoted on each image. Three different photon energies were used in the experiments as indicated in Table I.

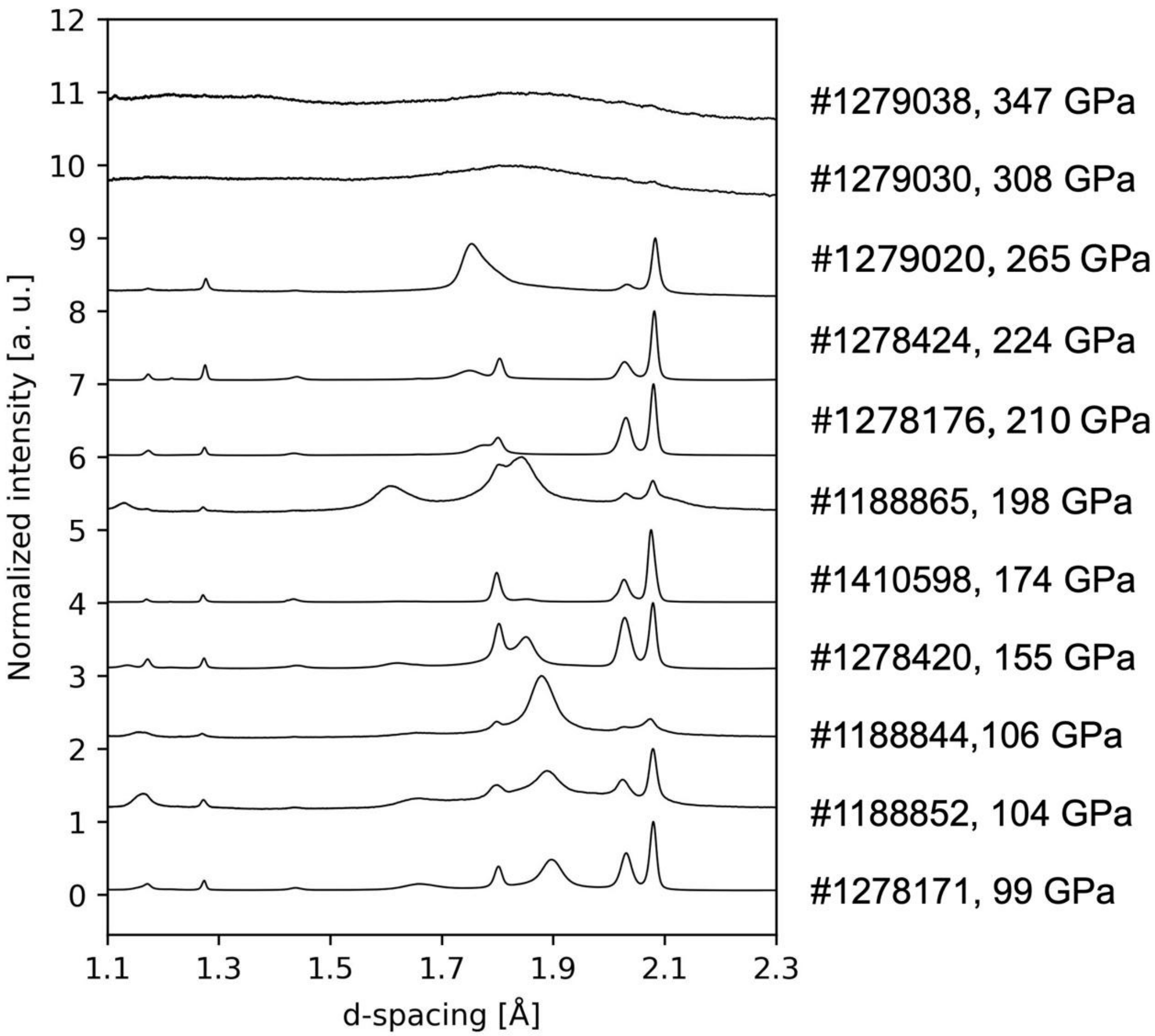

**Supplementary Fig. 5. Summary of the XRD line profiles for the shock experiments.** Each profile is normalized to the highest intensity peak, and an offset of 1 along the y-axis is applied for better visibility. Three different photon energies were used in the experiments as indicated in Supplementary Table I.

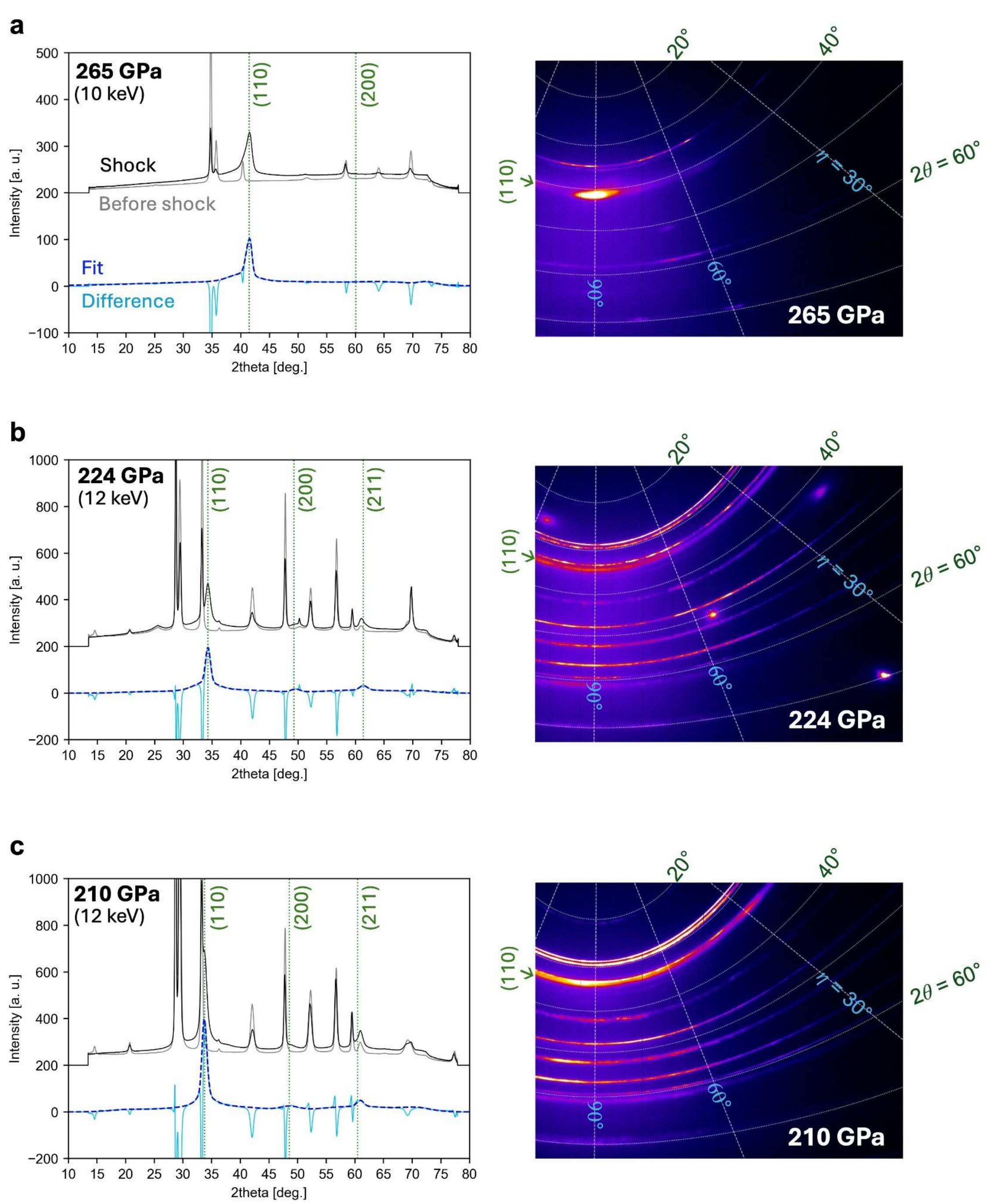

**Supplementary Fig. 6. Formation of the bcc phase.** The gray and black profile (both offset in y-axis for better visibility) are the before-shock and during-shock profiles. The shock pressures are (**a**) 265 ± 40 GPa, (**b**) 224 ± 30 GPa, and (**c**) 210 ± 29 GPa. The light-blue profile is obtained by subtracting the gray from black. The blue dashed profile is obtained by fitting the light-blue profile after manually masking the sharp uncompressed peaks. The green vertical lines indicate the predicted peak positions of bcc with lattice constants of $a$ = 2.478 Å, 2.480 Å, and 2.513 Å for a, b, and c, respectively. Noted numbers in parenthesis indicate the diffracting planes. The 265 ± 40 GPa shot (a) is measured using X-rays with a photon energy of $E$ = 10.00 keV ($\lambda$ = 1.240

Å), while 224 and 210 ± 29 GPa ones are measured at $E$ = 12.00 keV ($\lambda$ = 1.033 Å). The second peak for (a) is absent thought to be because of the texture developed which is implied from the spotty peak from the (110) diffractions.

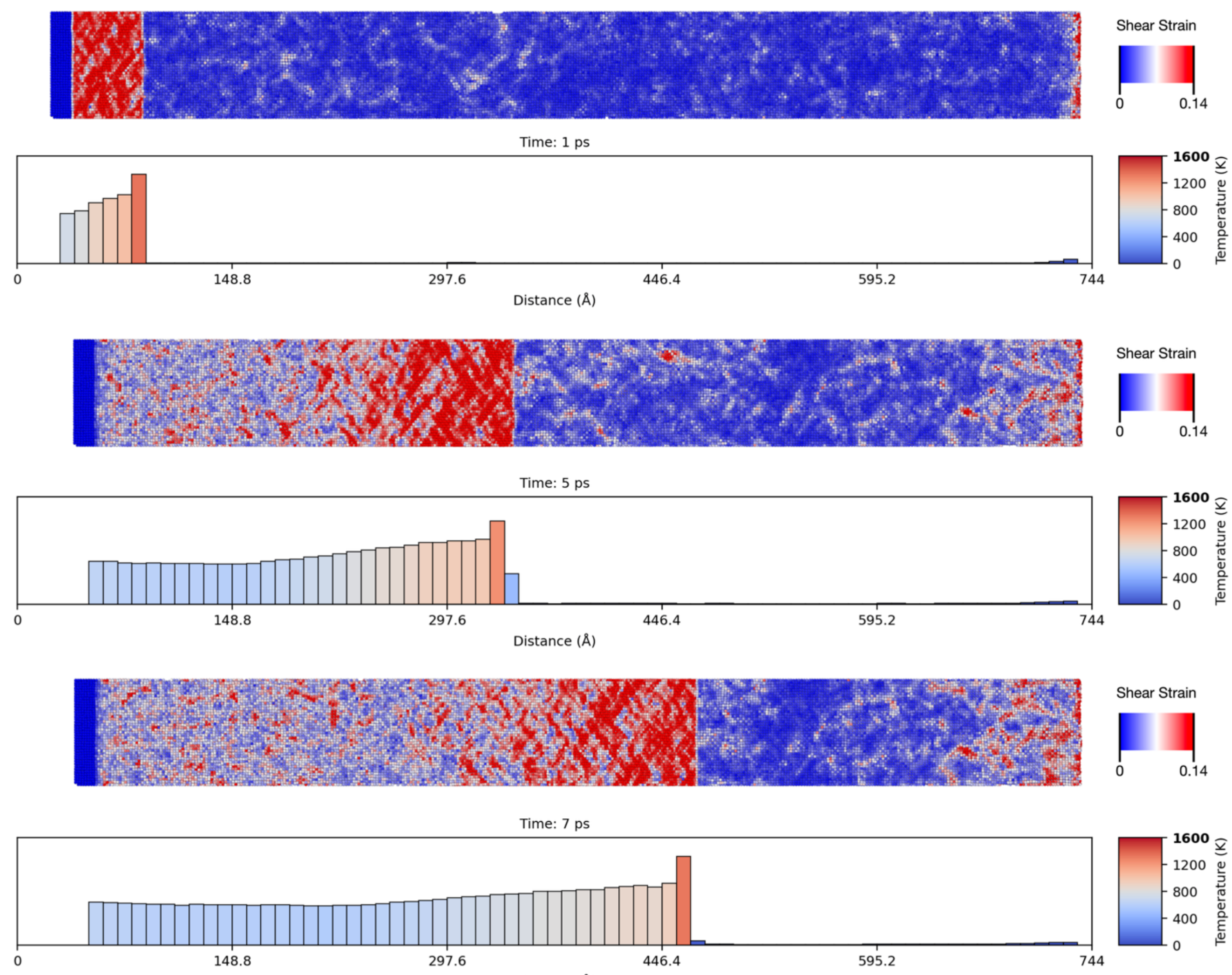

**Supplementary Fig. 7: MD simulations of strain and temperature as function of time**. A targeted MD simulation was performed with spatial binning of atomic velocities to compute the local instantaneous kinetic temperature, allowing us to resolve transient heating during uniaxial shock compression and to quantify its temporal correlation with the onset of plasticity.

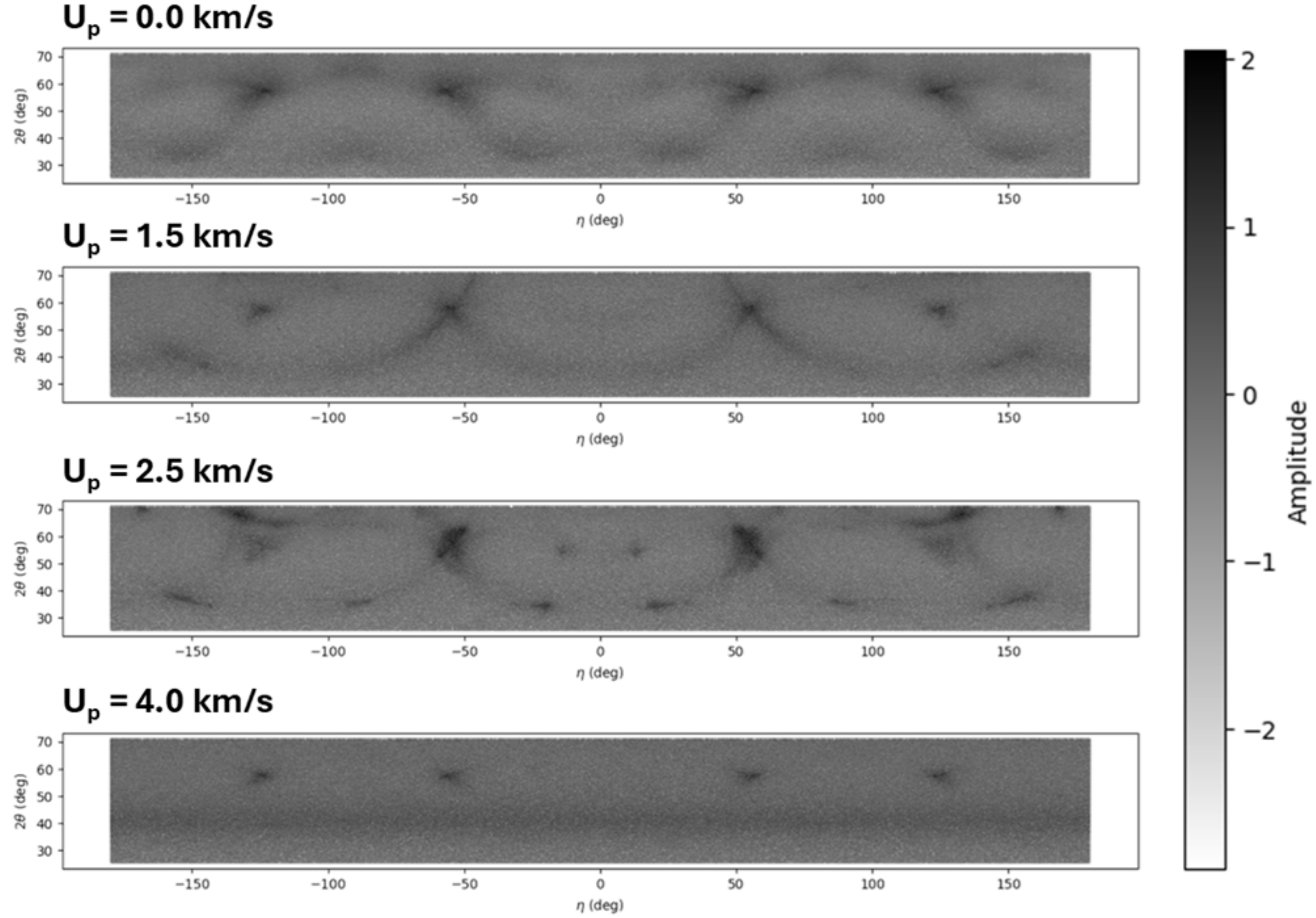

**Supplementary Fig. 8. Simulated X-ray diffraction patterns of the MD simulation results of shocked AM EHEA $AlCoCrFeNi_{2.1}$.** The shock velocities are (**a**) 0 km/s (**b**) 1.5 km/s (**c**) 2.5 km/s (**d**) 4 km/s. These simulations show the emergence of the amorphous phase in Fig 9 as a broad band and demonstrate the coexistence of bcc and amorphous phase at this velocity. The unshocked region is shown as 4 diffraction spots in the top of each image.